\documentclass[RNAAS]{aastex63}

\shorttitle{DQ Tau}
\shortauthors{Getman}

\graphicspath{{./}}

\begin{document}

\title{The Young Binary DQ Tau Produces Another X-ray Flare Near Periastron}

\correspondingauthor{Konstantin Getman}
\email{kug1@psu.edu}

\author[0000-0002-6137-8280]{Konstantin V. Getman}
\affiliation{Department of Astronomy \& Astrophysics \\
Pennsylvania State University \\ 
525 Davey Laboratory \\
University Park, PA 16802, USA}

\author[0000-0002-4324-3809]{Vitaly V. Akimkin}
\affiliation{Institute of Astronomy\\
Russian Academy of Sciences\\
48 Pyatnitskaya St., Moscow, 119017, Russia}

\author[0000-0003-2631-5265]{Nicole Arulanantham}
\affiliation{Space Telescope Science Institute\\ 3700 San Martin Drive, Baltimore, MD 21218, USA}

\author[0000-0001-7157-6275]{\'{A}gnes K\'{o}sp\'{a}l}
\affiliation{Konkoly Observatory, Research Centre for Astronomy and Earth Sciences, E\"otv\"os Lor\'and Research Network (ELKH), Konkoly-Thege Mikl\'os \'ut 15-17, 1121 Budapest, Hungary}
\affiliation{Max Planck Institute for Astronomy, K\"onigstuhl 17, 69117 Heidelberg, Germany}
\affiliation{ELTE E\"otv\"os Lor\'and University, Institute of Physics, P\'azm\'any P\'eter s\'et\'any 1/A, 1117 Budapest, Hungary}

\author[0000-0002-3913-7114]{Dmitry A. Semenov}
\affiliation{Max Planck Institute for Astronomy, K\"onigstuhl 17, 69117 Heidelberg, Germany}

\author[0000-0002-6911-8686]{Grigorii V. Smirnov-Pinchukov}
\affiliation{Max Planck Institute for Astronomy, K\"onigstuhl 17, 69117 Heidelberg, Germany}

\author[0000-0002-1284-5831]{Sierk E. van Terwisga}
\affiliation{Max Planck Institute for Astronomy, K\"onigstuhl 17, 69117 Heidelberg, Germany}

\begin{abstract}
This work is part of a multi-wavelength program to study the effects of X-ray/UV/optical stellar radiation on the chemistry of the circumbinary disk around the young high-eccentricity binary DQ~Tau. ALMA observations for near/around December 5, 2021 periastron were postponed due to bad weather, but supporting Swift-XRT-UVOT TOO observations were successful. These Swift observations along with previous X-ray-optical-mm data show that DQ Tau keeps exhibiting powerful flares near periastron, offering a unique laboratory for studies of flare effects on the gas-phase ion chemistry in protoplanetary disks. 
\end{abstract}

\section{Introduction}
X-ray flares are predicted to induce time-variable ion-molecular chemistry in protoplanetary disks around young stars \citep{Waggoner2022}. Variable H$^{13}$CO$^{+}$ emission has indeed been detected by ALMA in the disk around the young star IM Lup \citep{Cleeves2017}. However, no coincident X-ray observations were made and therefore the cause cannot be confidently attributed to X-ray flaring yet.

The occurrence of large X-ray flares in young stars, which are much more powerful and frequent than solar flares \citep{Getman2021}, generally cannot be predicted due to the stochastic nature of flares. Only in DQ Tau can the occurrence of large optical/mm/X-ray flux increases be reasonably predicted. DQ Tau is a $<3$~Myr old, nearby ($d=195$~pc), non-eclipsing, high eccentricity ($e=0.57$) binary, comprised of two $0.8$~M$_{\odot}$ stars with a robust orbital period of only 15.8~days \citep{Czekala2016, Fiorellino2022}. It is surrounded by a normal protoplanetary disk (with a small $0.3$~au cavity) of dust mass $\sim 70$~M$_{\oplus}$ and size $[25-100]$~au \citep{Czekala2016,Kospal2018,Ballering2019}.

At closest approach, the binary separation is uniquely small, only several to $10$ stellar radii. Optical and UV brightenings occur at orbital phase $\Phi = [0.8 - 1.2]$ and are attributed mainly to pulsed accretion of disk material onto the binary components \citep{Tofflemire2017, Kospal2018,Muzerolle2019,Fiorellino2022}. However, \textit{HST} spectra acquired at four phases $\left(\Phi = 0, 0.25, 0.5, 0.7 \right)$ show no correlations between orbital phase and NUV excess or C IV line emission generated at the accretion shock \citep{Ardila2015}, indicating that higher cadence observations are required to characterize the accretion behavior at UV wavelengths. DQ Tau exhibits large mm flares coincident with periastron passage \citep{Salter2008,Salter2010}. Prior to the current study, the only observation of the periastron passage with an X-ray telescope ({\it Chandra}) resulted in the detection of a large X-ray flare, coincident with a large mm IRAM flare \citep{Getman2011}. X-ray emission was attributed to collisions between the magnetospheres of the binary components, as suggested by the recurrence of synchrotron mm flaring in four periastron encounters, by the timing and energy relationships between the IRAM mm and {\it Chandra} X-ray flares, and by consistency between the flare loop size and binary separation \citep{Salter2010,Getman2011}. 

Our team was granted ALMA time to observe the reaction of H$^{13}$CO$^{+}$ emission across one orbit to an increase in ionizing radiation near/around related periastron passage. The original ALMA observations for near/around December 5, 2021 periastron were postponed due to bad weather. But supporting Swift-XRT-UVOT TOO observations were successful and are described below. 

\section{Analysis and Results}

Employing the Neil Gehrels Swift observatory \citep{Gehrels2004} we have conducted 15 short ($\sim 1.5$~ks) observations (separated by several hours) of DQ Tau near periastron over the December $[3-7]$, 2021 period (target-id: 14857). XRT was operating in the PC mode; UVOT - in the 0x30ed standard six-filter blue-weighted mode. X-ray lightcurve and source/background spectra along with related calibration files are constructed using the Swift-XRT data product generator \citep{Evans2007,Evans2009}, relied on the Heasoft package, version 6.29. All 15 low signal-to-noise source spectra were merged into a single spectrum. Using XSPEC \citep[part of Heasoft;][]{Arnaud1996}, this spectrum is fitted with two-temperature APEC plasma emission and TBABS X-ray absorption models \citep{Smith2001,Wilms00} keeping the cooler plasma temperature and column density fixed at $kT_1 = 0.7$~keV and  $N_H=1.3 \times 10^{21}$~cm$^{-2}$, respectively \citep{Getman2011}. The resulting time-averaged temperature of the hotter plasma component is $kT_2 \sim 2.6$~keV. The resulting time-averaged XRT count rate of 15.8~cnts/ksec and X-ray luminosity of $L_X=3.1 \times 10^{30}$~erg~s$^{-1}$ provide apparent-to-intrinsic X-ray flux conversion. Lightcurves for the six UVOT filters (V:B:U:W1:M2:W2) are constructed by applying the FAPPEND and UVOTMAGHIST tools from Heasoft to the Level II UVOT data. 

\begin{figure*}[ht!]
\epsscale{1.15}
\plotone{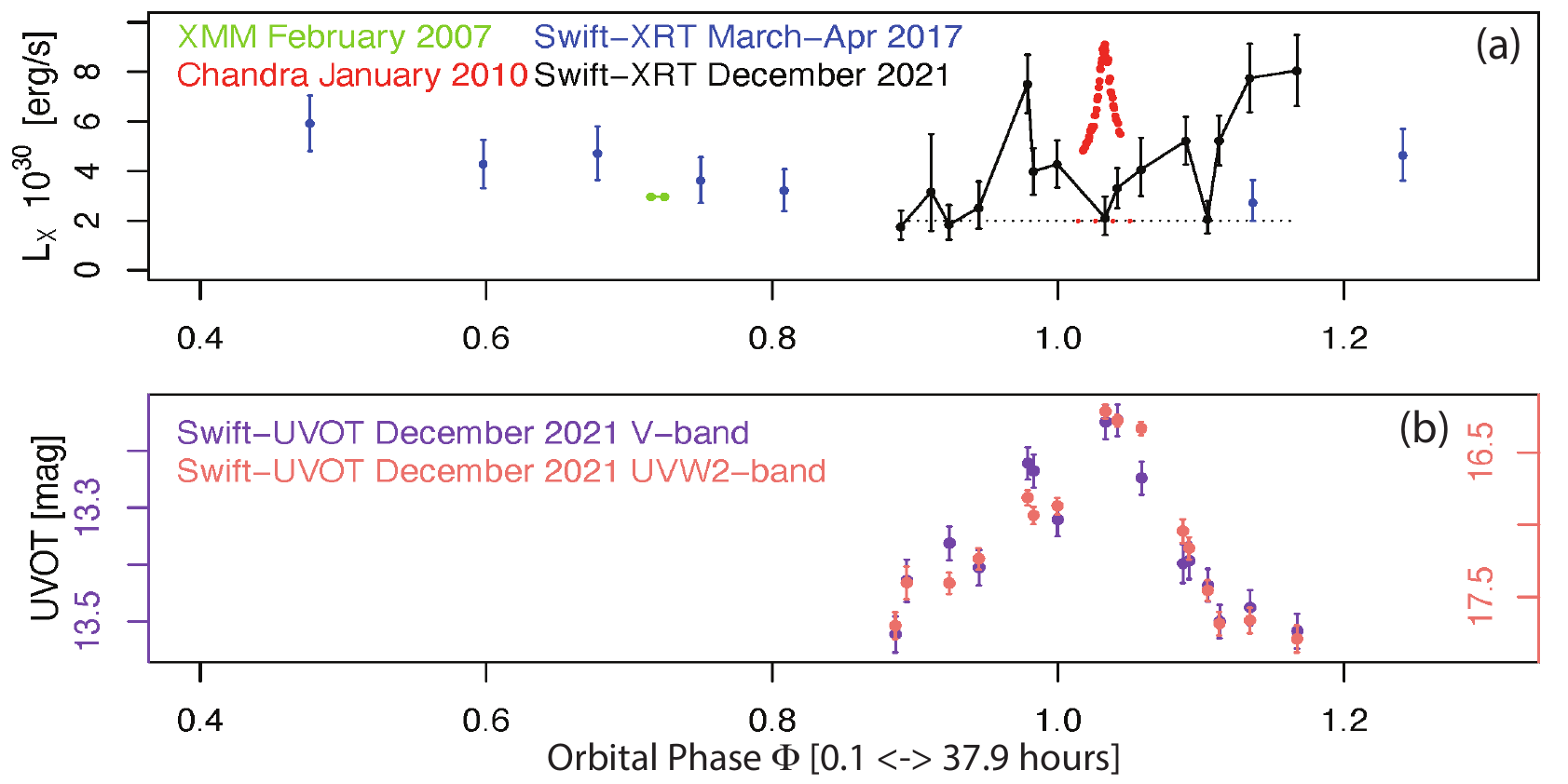}
\caption{Swift-XRT-UVOT lightcurves near December 05, 2021 periastron passage are compared to previous archived Swift, and {\it XMM}/{\it Chandra} X-ray \citep{Getman2011} data. X-ray ``characteristic'' level is marked by the dotted lines. X-ray luminosity is for the $[0.5 - 8]$~keV band assuming the new {\it Gaia}-based distance $d = 195$~pc. \label{fig1}}
\end{figure*}

Figure~\ref{fig1} shows the inferred XRT and UVOT lightcurves. Table~\ref{tab:tbl1} lists corresponding data. Figure~\ref{fig1}a also incorporates lightcurves for other data obtained with modern X-ray telescopes: Swift-2017 from archive, {\it XMM} and {\it Chandra} from \citet{Getman2011}. There are no signs of long-term variability --- the ``characteristic'' level of $L_X \sim 2 \times 10^{30}$~erg~s$^{-1}$ (i.e., superposition of numerous un-resolved flares) remains the same over the 11-year period as indicated by the black and red dotted lines in Figure~\ref{fig1}a. 

Remarkably, during the December 3--7, 2021 periastron passage DQ~Tau is found to exhibit an X-ray super-flare, whose morphology and energetics ($E_X \sim 4 \times 10^{35}$~erg) resemble those of the X-ray flare captured by {\it Chandra} during the January 2010 periastron passage. This December 2021 X-ray super-flare is followed by an unusually slow rise of X-ray luminosity (long-rise event; LRE) that is likely part of an even more complex and energetic ($E_X > 5 \times 10^{35}$~erg) flaring event; unfortunately, the full development of the event is truncated by the end of the Swift exposure. While large X-ray flares are detected near periastron (black and red), mildly elevated X-rays are seen away from periastron (blue and green).

During this December 3--7, 2021 periastron passage Swift-UVOT emission appears to be also variable, exhibiting at least two peaks --- the first peak is coincident in time with the peak of the X-ray super-flare; the second peak is near the beginning of the LRE event. Such solar-type optical-UV-X-ray flare peak co-occurrence exhibited by many young diskless stars \citep{Flaccomio2018} indicates contribution of chromospheric heating by magnetic reconnection to the accretion related optical brightening. The scenario of colliding magnetospheres triggering magnetic reconnection remains the most plausible explanation for the elevated X-ray emission near periastron. Due to such periodic X-ray flaring behaviour the disk-bearing system DQ Tau offers a unique laboratory for studies of flare effects on the gas-phase ion chemistry in protoplanetary disks.

A.K. acknowledges support from the European Research Council (ERC) under the European Union's Horizon 2020 research and innovation programme under grant agreement No 716155 (SACCRED). This work made use of data supplied by the UK Swift Science Data Centre at the University of Leicester.

\startlongtable
\begin{deluxetable*}{ccccc}
\tabletypesize{\small}
\tablecaption{X-ray And UVOT Data Shown In Figure~\ref{fig1}  \label{tab:tbl1}}
\tablewidth{0pt}
\tablehead{
\colhead{Instrument/Epoch} & \colhead{JD} &
\colhead{$\Phi$} & \colhead{$L_X$ or UVOT Magnitude} & \colhead{Error on $L_X$ or UVOT Magnitude}\\
\colhead{} & \colhead{(day)} &
\colhead{} & \colhead{($10^{30}$~erg~s$^{-1}$) or (mag)} & \colhead{($10^{30}$~erg~s$^{-1}$) or (mag)}\\
\colhead{(1)} & \colhead{(2)} & \colhead{(3)} & \colhead{(4)} & \colhead{(5)}
}
\startdata
\multicolumn{5}{c}{\bf (a) X-ray Data} \\
XMM/2007 & 2454144.680000 & 0.7153 & 2.97 & 0.42\\
XMM/2007 & 2454144.830000 & 0.7248 & 2.97 & 0.42\\
Chandra/2010 & 2455208.163231 & 1.0176 & 4.81 & 0.41\\
Chandra/2010 & 2455208.175963 & 1.0185 & 4.91 & 0.38\\
Chandra/2010 & 2455208.187421 & 1.0192 & 4.95 & 0.41\\
Chandra/2010 & 2455208.198995 & 1.0199 & 5.11 & 0.37\\
Chandra/2010 & 2455208.210569 & 1.0206 & 5.17 & 0.41\\
Chandra/2010 & 2455208.222722 & 1.0214 & 5.23 & 0.40\\
Chandra/2010 & 2455208.232676 & 1.0220 & 5.39 & 0.40\\
Chandra/2010 & 2455208.243208 & 1.0227 & 5.62 & 0.43\\
Chandra/2010 & 2455208.253741 & 1.0234 & 5.62 & 0.43\\
Chandra/2010 & 2455208.263926 & 1.0240 & 5.80 & 0.42\\
Chandra/2010 & 2455208.274342 & 1.0247 & 5.74 & 0.45\\
Chandra/2010 & 2455208.284759 & 1.0253 & 5.80 & 0.43\\
Chandra/2010 & 2455208.293555 & 1.0259 & 6.24 & 0.42\\
Chandra/2010 & 2455208.302930 & 1.0265 & 6.24 & 0.39\\
Chandra/2010 & 2455208.311727 & 1.0270 & 6.47 & 0.43\\
Chandra/2010 & 2455208.320754 & 1.0276 & 6.87 & 0.40\\
Chandra/2010 & 2455208.329319 & 1.0282 & 6.99 & 0.40\\
Chandra/2010 & 2455208.336727 & 1.0286 & 7.37 & 0.43\\
Chandra/2010 & 2455208.344366 & 1.0291 & 7.62 & 0.46\\
Chandra/2010 & 2455208.352236 & 1.0296 & 8.12 & 0.40\\
Chandra/2010 & 2455208.359991 & 1.0301 & 8.10 & 0.46\\
Chandra/2010 & 2455208.367282 & 1.0306 & 8.36 & 0.50\\
Chandra/2010 & 2455208.374574 & 1.0310 & 8.49 & 0.49\\
Chandra/2010 & 2455208.381171 & 1.0314 & 8.91 & 0.52\\
Chandra/2010 & 2455208.388463 & 1.0319 & 8.75 & 0.47\\
Chandra/2010 & 2455208.395986 & 1.0324 & 8.99 & 0.50\\
Chandra/2010 & 2455208.403741 & 1.0329 & 9.11 & 0.47\\
Chandra/2010 & 2455208.410801 & 1.0333 & 8.85 & 0.53\\
Chandra/2010 & 2455208.418671 & 1.0338 & 8.57 & 0.53\\
Chandra/2010 & 2455208.426657 & 1.0343 & 8.47 & 0.50\\
Chandra/2010 & 2455208.434296 & 1.0348 & 8.38 & 0.44\\
Chandra/2010 & 2455208.442282 & 1.0353 & 8.49 & 0.52\\
Chandra/2010 & 2455208.451194 & 1.0359 & 7.62 & 0.47\\
Chandra/2010 & 2455208.459759 & 1.0364 & 7.74 & 0.49\\
Chandra/2010 & 2455208.469597 & 1.0370 & 7.19 & 0.47\\
Chandra/2010 & 2455208.478625 & 1.0376 & 7.11 & 0.44\\
Chandra/2010 & 2455208.487305 & 1.0382 & 6.95 & 0.44\\
Chandra/2010 & 2455208.497722 & 1.0388 & 6.63 & 0.41\\
Chandra/2010 & 2455208.508602 & 1.0395 & 6.53 & 0.39\\
Chandra/2010 & 2455208.519250 & 1.0402 & 6.22 & 0.43\\
Chandra/2010 & 2455208.529551 & 1.0408 & 6.08 & 0.36\\
Chandra/2010 & 2455208.539389 & 1.0415 & 6.02 & 0.43\\
Chandra/2010 & 2455208.551310 & 1.0422 & 5.92 & 0.44\\
Chandra/2010 & 2455208.562305 & 1.0429 & 5.58 & 0.38\\
Chandra/2010 & 2455208.574342 & 1.0437 & 5.48 & 0.41\\
Swift/2017 & 2457822.679464 & 0.4768 & 5.93 & 1.12\\
Swift/2017 & 2457824.594431 & 0.5980 & 4.29 & 0.98\\
Swift/2017 & 2457825.856935 & 0.6779 & 4.72 & 1.08\\
Swift/2017 & 2457826.998317 & 0.7501 & 3.63 & 0.93\\
Swift/2017 & 2457827.921509 & 0.8086 & 3.22 & 0.84\\
Swift/2017 & 2457833.100874 & 1.1363 & 2.73 & 0.83\\
Swift/2017 & 2457834.758614 & 1.2412 & 4.64 & 1.04\\
Swift/2017 & 2457836.358990 & 1.3425 & 4.09 & 0.96\\
Swift/2017 & 2457836.629495 & 1.3596 & 2.35 & 0.72\\
Swift/2017 & 2457847.113933 & 2.0231 & 5.54 & 1.36\\
Swift/2017 & 2457848.377409 & 2.1031 & 3.24 & 0.83\\
Swift/2017 & 2457850.037880 & 2.2082 & 4.42 & 1.06\\
Swift/2017 & 2457850.768441 & 2.2544 & 2.49 & 0.79\\
Swift/2021 & 2459551.579577 & 0.8899 & 1.75 & 0.59\\
Swift/2021 & 2459551.915829 & 0.9112 & 3.16 & 1.96\\
Swift/2021 & 2459552.113757 & 0.9237 & 1.84 & 0.70\\
Swift/2021 & 2459552.442115 & 0.9445 & 2.51 & 0.95\\
Swift/2021 & 2459552.981284 & 0.9786 & 7.51 & 1.18\\
Swift/2021 & 2459553.046312 & 0.9828 & 3.99 & 0.94\\
Swift/2021 & 2459553.310251 & 0.9995 & 4.28 & 0.96\\
Swift/2021 & 2459553.839425 & 1.0329 & 2.10 & 0.78\\
Swift/2021 & 2459553.972205 & 1.0413 & 3.31 & 0.81\\
Swift/2021 & 2459554.236366 & 1.0581 & 4.06 & 1.18\\
Swift/2021 & 2459554.731282 & 1.0894 & 5.22 & 0.96\\
Swift/2021 & 2459554.971614 & 1.1046 & 2.07 & 0.65\\
Swift/2021 & 2459555.099986 & 1.1127 & 5.23 & 1.00\\
Swift/2021 & 2459555.437964 & 1.1341 & 7.75 & 1.38\\
Swift/2021 & 2459555.958085 & 1.1670 & 8.05 & 1.44\\
\multicolumn{5}{c}{\bf (b) Swift-UVOT 2021 Data} \\
V & 2459551.520480 & 0.8862 & 13.523 & 0.031\\
V & 2459551.645673 & 0.8941 & 13.428 & 0.037\\
V & 2459552.117767 & 0.9240 & 13.362 & 0.030\\
V & 2459552.443384 & 0.9446 & 13.405 & 0.031\\
V & 2459552.984216 & 0.9788 & 13.222 & 0.028\\
V & 2459553.048982 & 0.9829 & 13.236 & 0.029\\
V & 2459553.313229 & 0.9996 & 13.321 & 0.029\\
V & 2459553.842633 & 1.0331 & 13.149 & 0.030\\
V & 2459553.975671 & 1.0416 & 13.147 & 0.028\\
V & 2459554.239237 & 1.0582 & 13.248 & 0.029\\
V & 2459554.698611 & 1.0873 & 13.398 & 0.034\\
V & 2459554.767455 & 1.0917 & 13.394 & 0.032\\
V & 2459554.974599 & 1.1048 & 13.436 & 0.029\\
V & 2459555.103296 & 1.1129 & 13.500 & 0.029\\
V & 2459555.440189 & 1.1342 & 13.476 & 0.031\\
V & 2459555.960720 & 1.1672 & 13.517 & 0.031\\
UVW2 & 2459551.518152 & 0.8860 & 17.699 & 0.093\\
UVW2 & 2459551.644507 & 0.8940 & 17.403 & 0.113\\
UVW2 & 2459552.115006 & 0.9238 & 17.405 & 0.074\\
UVW2 & 2459552.441058 & 0.9444 & 17.236 & 0.074\\
UVW2 & 2459552.980883 & 0.9786 & 16.814 & 0.053\\
UVW2 & 2459553.046117 & 0.9827 & 16.936 & 0.060\\
UVW2 & 2459553.310218 & 0.9995 & 16.870 & 0.057\\
UVW2 & 2459553.840293 & 1.0330 & 16.217 & 0.049\\
UVW2 & 2459553.972136 & 1.0413 & 16.280 & 0.042\\
UVW2 & 2459554.236318 & 1.0581 & 16.334 & 0.046\\
UVW2 & 2459554.696999 & 1.0872 & 17.044 & 0.081\\
UVW2 & 2459554.765409 & 1.0915 & 17.164 & 0.077\\
UVW2 & 2459554.971244 & 1.1046 & 17.456 & 0.071\\
UVW2 & 2459555.100086 & 1.1127 & 17.685 & 0.081\\
UVW2 & 2459555.437561 & 1.1341 & 17.662 & 0.089\\
UVW2 & 2459555.958097 & 1.1670 & 17.791 & 0.093\\
\enddata
\tablecomments{This table consists of two parts. Part (a) presents all available X-ray data obtained by modern telescopes ($XMM$, $Chandra$, $Swift$), including our recent {\it Swift}-XRT 2021 data. Part (b) gives our recent {\it Swift}-UVOT 2021 data for two filters, V and W2.   Column 1: Instrument and epoch in Part (a). Type of UVOT filter in Part (b). Columns 2-3: JD epoch and corresponding orbital phase. Columns 4-5: X-ray luminosity in the $(0.5-8)$~keV band and its 68\% error (Part (a)); magnitude and its 68\% error (Part (b)). The {\it Chandra} (red) and {\it XMM} (green) data from \citet{Getman2011} were corrected for the new {\it Gaia}-based distance ($d=195$~pc) and new orbital solution \citep{Czekala2016,Fiorellino2022}.}
\end{deluxetable*}

\bibliography{my_bibliography}{}

\begin{thebibliography}{}
\expandafter\ifx\csname natexlab\endcsname\relax\def\natexlab#1{#1}\fi
\providecommand{\url}[1]{\href{#1}{#1}}
\providecommand{\dodoi}[1]{doi:~\href{http://doi.org/#1}{\nolinkurl{#1}}}
\providecommand{\doeprint}[1]{\href{http://ascl.net/#1}{\nolinkurl{http://ascl.net/#1}}}
\providecommand{\doarXiv}[1]{\href{https://arxiv.org/abs/#1}{\nolinkurl{https://arxiv.org/abs/#1}}}

\bibitem[{{Ardila} {et~al.}(2015){Ardila}, {Jonhs-Krull}, {Herczeg}, {Mathieu},
  \& {Quijano-Vodniza}}]{Ardila2015}
{Ardila}, D.~R., {Jonhs-Krull}, C., {Herczeg}, G.~J., {Mathieu}, R.~D., \&
  {Quijano-Vodniza}, A. 2015, \apj, 811, 131,
  \dodoi{10.1088/0004-637X/811/2/131}

\bibitem[{{Arnaud}(1996)}]{Arnaud1996}
{Arnaud}, K.~A. 1996, in Astronomical Society of the Pacific Conference Series,
  Vol. 101, Astronomical Data Analysis Software and Systems V, ed. G.~H.
  {Jacoby} \& J.~{Barnes}, 17

\bibitem[{{Ballering} \& {Eisner}(2019)}]{Ballering2019}
{Ballering}, N.~P., \& {Eisner}, J.~A. 2019, \aj, 157, 144,
  \dodoi{10.3847/1538-3881/ab0a56}

\bibitem[{{Cleeves} {et~al.}(2017){Cleeves}, {Bergin}, {{\"O}berg}, {Andrews},
  {Wilner}, \& {Loomis}}]{Cleeves2017}
{Cleeves}, L.~I., {Bergin}, E.~A., {{\"O}berg}, K.~I., {et~al.} 2017, \apjl,
  843, L3, \dodoi{10.3847/2041-8213/aa76e2}

\bibitem[{{Czekala} {et~al.}(2016){Czekala}, {Andrews}, {Torres}, {Jensen},
  {Stassun}, {Wilner}, \& {Latham}}]{Czekala2016}
{Czekala}, I., {Andrews}, S.~M., {Torres}, G., {et~al.} 2016, \apj, 818, 156,
  \dodoi{10.3847/0004-637X/818/2/156}

\bibitem[{{Evans} {et~al.}(2007){Evans}, {Beardmore}, {Page}, {Tyler},
  {Osborne}, {Goad}, {O'Brien}, {Vetere}, {Racusin}, {Morris}, {Burrows},
  {Capalbi}, {Perri}, {Gehrels}, \& {Romano}}]{Evans2007}
{Evans}, P.~A., {Beardmore}, A.~P., {Page}, K.~L., {et~al.} 2007, \aap, 469,
  379, \dodoi{10.1051/0004-6361:20077530}

\bibitem[{{Evans} {et~al.}(2009){Evans}, {Beardmore}, {Page}, {Osborne},
  {O'Brien}, {Willingale}, {Starling}, {Burrows}, {Godet}, {Vetere}, {Racusin},
  {Goad}, {Wiersema}, {Angelini}, {Capalbi}, {Chincarini}, {Gehrels}, {Kennea},
  {Margutti}, {Morris}, {Mountford}, {Pagani}, {Perri}, {Romano}, \&
  {Tanvir}}]{Evans2009}
---. 2009, \mnras, 397, 1177, \dodoi{10.1111/j.1365-2966.2009.14913.x}

\bibitem[{{Fiorellino} {et~al.}(2022){Fiorellino}, {Park}, {K{\'o}sp{\'a}l}, \&
  {{\'A}brah{\'a}m}}]{Fiorellino2022}
{Fiorellino}, E., {Park}, S., {K{\'o}sp{\'a}l}, {\'A}., \& {{\'A}brah{\'a}m},
  P. 2022, arXiv e-prints, arXiv:2201.00784.
\newblock \doarXiv{2201.00784}

\bibitem[{{Flaccomio} {et~al.}(2018){Flaccomio}, {Micela}, {Sciortino}, {Cody},
  {Guarcello}, {Morales-Calder{\`o}n}, {Rebull}, \& {Stauffer}}]{Flaccomio2018}
{Flaccomio}, E., {Micela}, G., {Sciortino}, S., {et~al.} 2018, \aap, 620, A55,
  \dodoi{10.1051/0004-6361/201833308}

\bibitem[{{Gehrels} {et~al.}(2004){Gehrels}, {Chincarini}, {Giommi}, {Mason},
  {Nousek}, {Wells}, {White}, {Barthelmy}, {Burrows}, {Cominsky}, {Hurley},
  {Marshall}, {M{\'e}sz{\'a}ros}, {Roming}, {Angelini}, {Barbier}, {Belloni},
  {Campana}, {Caraveo}, {Chester}, {Citterio}, {Cline}, {Cropper}, {Cummings},
  {Dean}, {Feigelson}, {Fenimore}, {Frail}, {Fruchter}, {Garmire}, {Gendreau},
  {Ghisellini}, {Greiner}, {Hill}, {Hunsberger}, {Krimm}, {Kulkarni}, {Kumar},
  {Lebrun}, {Lloyd-Ronning}, {Markwardt}, {Mattson}, {Mushotzky}, {Norris},
  {Osborne}, {Paczynski}, {Palmer}, {Park}, {Parsons}, {Paul}, {Rees},
  {Reynolds}, {Rhoads}, {Sasseen}, {Schaefer}, {Short}, {Smale}, {Smith},
  {Stella}, {Tagliaferri}, {Takahashi}, {Tashiro}, {Townsley}, {Tueller},
  {Turner}, {Vietri}, {Voges}, {Ward}, {Willingale}, {Zerbi}, \&
  {Zhang}}]{Gehrels2004}
{Gehrels}, N., {Chincarini}, G., {Giommi}, P., {et~al.} 2004, \apj, 611, 1005,
  \dodoi{10.1086/422091}

\bibitem[{{Getman} {et~al.}(2011){Getman}, {Broos}, {Salter}, {Garmire}, \&
  {Hogerheijde}}]{Getman2011}
{Getman}, K.~V., {Broos}, P.~S., {Salter}, D.~M., {Garmire}, G.~P., \&
  {Hogerheijde}, M.~R. 2011, \apj, 730, 6, \dodoi{10.1088/0004-637X/730/1/6}

\bibitem[{{Getman} \& {Feigelson}(2021)}]{Getman2021}
{Getman}, K.~V., \& {Feigelson}, E.~D. 2021, \apj, 916, 32,
  \dodoi{10.3847/1538-4357/ac00be}

\bibitem[{{K{\'o}sp{\'a}l} {et~al.}(2018){K{\'o}sp{\'a}l}, {{\'A}brah{\'a}m},
  {Zsidi}, {Vida}, {Szab{\'o}}, {Mo{\'o}r}, \& {P{\'a}l}}]{Kospal2018}
{K{\'o}sp{\'a}l}, {\'A}., {{\'A}brah{\'a}m}, P., {Zsidi}, G., {et~al.} 2018,
  \apj, 862, 44, \dodoi{10.3847/1538-4357/aacafa}

\bibitem[{{Muzerolle} {et~al.}(2019){Muzerolle}, {Flaherty}, {Balog}, {Beck},
  \& {Gutermuth}}]{Muzerolle2019}
{Muzerolle}, J., {Flaherty}, K., {Balog}, Z., {Beck}, T., \& {Gutermuth}, R.
  2019, \apj, 877, 29, \dodoi{10.3847/1538-4357/ab1756}

\bibitem[{{Salter} {et~al.}(2008){Salter}, {Hogerheijde}, \&
  {Blake}}]{Salter2008}
{Salter}, D.~M., {Hogerheijde}, M.~R., \& {Blake}, G.~A. 2008, \aap, 492, L21,
  \dodoi{10.1051/0004-6361:200810807}

\bibitem[{{Salter} {et~al.}(2010){Salter}, {K{\'o}sp{\'a}l}, {Getman},
  {Hogerheijde}, {van Kempen}, {Carpenter}, {Blake}, \& {Wilner}}]{Salter2010}
{Salter}, D.~M., {K{\'o}sp{\'a}l}, {\'A}., {Getman}, K.~V., {et~al.} 2010,
  \aap, 521, A32, \dodoi{10.1051/0004-6361/201015197}

\bibitem[{{Smith} {et~al.}(2001){Smith}, {Brickhouse}, {Liedahl}, \&
  {Raymond}}]{Smith2001}
{Smith}, R.~K., {Brickhouse}, N.~S., {Liedahl}, D.~A., \& {Raymond}, J.~C.
  2001, \apjl, 556, L91, \dodoi{10.1086/322992}

\bibitem[{{Tofflemire} {et~al.}(2017){Tofflemire}, {Mathieu}, {Ardila},
  {Akeson}, {Ciardi}, {Johns-Krull}, {Herczeg}, \&
  {Quijano-Vodniza}}]{Tofflemire2017}
{Tofflemire}, B.~M., {Mathieu}, R.~D., {Ardila}, D.~R., {et~al.} 2017, \apj,
  835, 8, \dodoi{10.3847/1538-4357/835/1/8}

\bibitem[{{Waggoner} \& {Cleeves}(2022)}]{Waggoner2022}
{Waggoner}, A.~R., \& {Cleeves}, L.~I. 2022, arXiv e-prints, arXiv:2202.06962.
\newblock \doarXiv{2202.06962}

\bibitem[{{Wilms} {et~al.}(2000){Wilms}, {Allen}, \& {McCray}}]{Wilms00}
{Wilms}, J., {Allen}, A., \& {McCray}, R. 2000, \apj, 542, 914,
  \dodoi{10.1086/317016}

\end{thebibliography}
\bibliographystyle{aasjournal}
\end{document}